# Amygdala and cortical gamma-band responses to emotional faces depend on the attended to valence

Enya M. Weidner[1], Stephan Moratti[2], Sebastian Schindler[3], Philip Grewe[4,5], Christian G. Bien[4], Johanna Kissler[1]

[1] Department of Psychology, Bielefeld University, Universitätsstraße 25, 33615 Bielefeld, Germany

[2] Department of Experimental Psychology, Complutense University of Madrid, Spain

[3] Institute of Medical Psychology and Systems Neuroscience, University of Münster, Von-Esmarch-Str. 52, 48149 Münster, Germany

[4] Department of Epileptology (Krankenhaus Mara), Bielefeld University, Medical School OWL, Maraweg 21, 33617 Bielefeld, Germany

[5] Clinical Neuropsychology and Epilepsy Research, Medical School OWL, Bielefeld University, Bielefeld, Germany

**ABSTRACT**

The amygdala is assumed to contribute to a bottom-up attentional bias during visual processing of emotional faces. Still, how its response to emotion interacts with top-down attention is not fully understood. It is also unclear to what extent amygdala activity and scalp EEG respond to emotion and attention in a similar way. Therefore, we studied the interaction of emotion and attention during face processing in oscillatory gamma-band activity (GBA, >35 Hz) in the amygdala and on the scalp. Amygdala signals were recorded via intracranial EEG (iEEG) recordings in 10 patients (6 right, 4 left) with epilepsy. Scalp recordings were collected from 19 healthy participants. Three randomized blocks of angry, neutral, and happy faces were presented, and either the negative, the neutral, or the positive expression was denoted as the target category. Both groups detected happy faces fastest and most accurately. During attention to negative faces, low GBA (< 90 Hz) increased specifically for angry faces both in the amygdala and over posterior scalp regions, albeit earlier on the scalp (60 ms) than in the amygdala (130 ms). From 220 ms, amygdala high GBA (117.5-145 Hz) was additionally persistently increased for both angry and neutral compared to happy faces. When neutral faces served as targets, amygdala high GBA (105-122.5 Hz) was higher for emotional than neutral faces from 160-320 ms. Attention to positive faces did not result in a differentiation between facial expressions. Present data reveal that attention-independent emotion detection in amygdala high GBA may only occur during a neutral focus of attention. Top-down threat vigilance coordinates widespread GBA, biasing stimulus processing in favor of negative faces. These results are in line with a multi-pathway model of emotion processing and help specify the role of GBA in this process by revealing how attentional focus can tune the timing and amplitude of emotional GBA responses.

**Impact Statement:** Amygdala and cortical gamma-band responses to emotional faces are tuned by the attentional focus. Attention to potential threats enhances widespread gamma in favor of angry faces. This implies top-down attentional guidance during threat vigilance. During a neutral focus, automatic attentional shifts to emotion occurred only in the amygdala, suggesting an involvement in the allocation of attentional resources towards unattended emotion independently of cortical signals.

Attention tunes gamma-band responses to faces

**INTRODUCTION**

Faces carry high biological significance, enhancing their visual processing compared to other objects. This is particularly true for faces expressing emotions (for review see Palermo & Rhodes, 2007; Schindler & Bublatzky, 2020). The ensuing bias is termed *emotional attention* (Vuilleumier, 2005; Vuilleumier & Huang, 2009) or *motivated attention* (Lang et al., 1997), and it is proposed to ensure enhanced processing of emotional material even without explicitly directed attention.

Face recognition involves increased cortical gamma-band activity (GBA), that is high frequency oscillations above 30 Hz, as well as gamma-synchronization in a distributed neural network that coordinates bottom-up and top-down processing (Lachaux et al., 2005; Moratti et al., 2014; for review see Güntekin & Başar, 2014). According to Jensen et al. (2012; 2014), it reflects computational processes that determine stimulus salience and enhanced GBA synchronization may be an indicator of attentional selection. Therefore, cortical GBA increases by emotional expressions, often localized in the visual cortex with magnetoencephalogram (MEG) source reconstructions (Luo et al., 2009; Müsch et al., 2017), corroborate the idea of emotional attention (Balconi & Lucchiari, 2008; Balconi & Pozzoli, 2009; Keil et al., 1999; Luo et al., 2009; Müsch et al., 2017).

Intracranial EEG (iEEG) (Guex et al., 2020; Sato et al., 2011) and MEG (Liu et al., 2015; Luo et al., 2007; Luo et al., 2009) studies show that amygdala GBA also favors emotional over neutral faces, particularly fast gamma (> 90 Hz) (2022; Guex et al., 2020). Emotion-driven enhancements of amygdala signals repeatedly preceded such enhancements in the visual cortex in iEEG (Méndez-Bértolo et al., 2016), MEG (Luo et al., 2007), and fast-sampled hemodynamic activity (Sabatinelli et al., 2009). Therefore, the amygdala might directly contribute to a visual processing bias in favor of emotional faces, supported by its numerous interconnections to sensory and associative areas (Amaral et al., 2003; Vuilleumier, 2005). Models of emotion processing propose that the amygdala evaluates stimulus relevance based on rapid (< 100-200 ms), rudimentary perceptual cues, either via a feedback loop with the visual cortex (Vuilleumier, 2005) or through subcortical connections that bypass cortical processing (LeDoux, 2007; Méndez-Bértolo et al., 2016). These early stages of emotional processing might be independent of higher cognitive control and top-down attention, as supported by studies on scalp event-related potentials (ERPs). Here, top-down attention usually drives emotional attention in late (~400-700 ms) but not in early to mid-latency (~200 ms) visual processing (Schindler & Kissler, 2016; Schindler et al., 2020; Schupp et al., 2007; for review see Schindler & Bublatzky, 2020).

Interestingly, intracranial event-related potentials (iERPs) show that top-down attention might affect amygdala iERPs to emotional faces already at 150-300 ms, earlier than top-down effects in scalp ERPs (Guex et al., 2020; Krolak-Salmon et al., 2004; Weidner et al., 2022). Thus, to sustain a behavioral goal, top-down attention, likely guided by pre-frontal areas (Ishii et al., 1999; Small et al., 2003; Ptak, 2012), might direct or even suppress amygdala activity (Marek et al., 2013; Pessoa et al., 2002; Pessoa et al., 2005; for review see Kaldewaij et al., 2016). However, ERPs and GBA reflect different neural dynamics (Bartolo et al., 2011; Cleeren et al., 2020; Engell & McCarthy, 2010) that potentially diverge in their response to top-down control but research about the interactions between emotional and top-down attention in GBA is still sparse.

Investigating seven patients with stereotactic amygdala implantations, Guex and colleagues (2020) directed participants' attention towards fearful or neutral faces during a serial face presentation. The detection of a target, either fearful or neutral, should be indicated by a button press. And indeed, attentional modulation of fear processing was present in amygdala iERPs at ~ 130 ms, whereas an early (~ 80 ms), attention-independent differentiation of fearful from neutral faces arose in high GBA (> 100 Hz). In contrast to iERPs, amygdala GBA may rapidly indicate fear-relevance, irrespective of the attentional context. This aligns with MEG data showing early (~ 40 ms), pre-attentive amygdala GBA increases to fearful faces (Luo et al., 2007; Luo et al., 2010).

Recent iEEG insights by Guex et al. (2022) show that amygdala GBA is responsive not only to fearful, but also to angry and happy expressions depending on eye gaze: GBA increased for angry and happy faces with directed relative to averted gaze, while for fearful faces increases were stronger with averted relative to directed gaze, similar to earlier fMRI studies (N'Diaye et al., 2009). Rather than exhibiting fear-specificity per se, this points towards a potential role of the amygdala in guiding attention based on social cues derived from several different emotional expressions (Sander et al., 2003). The question arises whether this attentional guidance occurs generally irrespective of higher attentional control. This might be true for fearful faces (Guex et al., 2022), but we still know little about amygdala GBA in response to other facial expressions and how it might be modulated depending on which expression is task-relevant. Some MEG studies already suggest that rapid, pre-attentive amygdala GBA increases can also occur for angry and happy faces (Liu et al., 2015), but this remains to be thoroughly addressed with the temporal and spatial precision of iEEG.

IEEG data by Müsch et al. (2014) show that top-down attention enhances amygdala GBA to faces at intermediate processing stages (~350 ms) when happy faces rather than a red-tint are monitored during a stream of fearful, neutral, and non-faces. But this increase was not emotion-specific, it occurred for both fearful and neutral faces. Since only happy faces were instructed as task-relevant, it remains unclear whether results represent mechanisms of face processing in general or might change if different facial expressions were instructed as attentionally relevant.

**The present study**

Our present goal is to further specify the interactions of emotion and attention during face processing in the amygdala and compare it to scalp GBA. We aim to extend evidence from fearful to other facial expressions. Our previous study showed that attentional modulations of amygdala iERPs in response to angry, neutral, and happy faces arise earlier than such modulations on the scalp (Weidner et al., 2022). We now complement these findings by investigating induced GBA in the amygdala, obtained from ten patients undergoing pre-surgical iEEG evaluation, and in scalp signals, obtained from healthy students. This enabled us to compare GBA directly from the amygdala to effects found in scalp EEG. During the task, we directed the participants' attention toward negative (angry), neutral, or positive (happy) faces. Because previous studies suggest a functional differentiation of amygdala low (< 90 Hz) and high GBA (> 90 Hz) (Guex et al., 2020; Guex et al., 2022), we analyzed low and high gamma frequencies separately. We anticipate a differentiation of emotional from neutral faces at ~150-300 ms, with the strongest GBA increase for angry faces, both in the amygdala and on the scalp (Balconi &

Lucchiari, 2008; Guex et al., 2020; Luo et al., 2007). We test to what extent these effects are further modified by the attentional focus. Given the paucity and inconsistency of previous data on this particular issue, we have no clear expectations regarding the direction and timing of such interactive effects.

## METHOD

**Sample**

*Patient sample.* The initial sample consisted of thirteen patients with unilateral or bilateral amygdala implantations. Selection criteria for analysis were first, accurate task performance, second, no severe anterior temporal lobe lesions, and third, no (subclinical) seizures during the experiment. One patient was excluded due to poor task performance (confusion of the target conditions). Two patients were excluded because of severe temporal lobe damage, leaving ten patients for the final sample (30% female, age: M = 36.70 years, range = 19-63). Clinical details of the patients remaining in the iEEG analysis can be found in Table 1. On average, the Beck Depression Inventory (BDI, Beck et al., 1996) showed signs of moderate depressive symptoms in the seven patients for whom these data were available (M = 23.50, SD = 6.188). Two of these patients fulfilled the criteria for severe depression (BDI Score: 31). No other clinically ascertained psychiatric diagnoses were present.

The patients were hospitalized at the Mara, Dept. of Epileptology of Bielefeld University in Bielefeld, Germany. They were undergoing pre-surgical iEEG assessment for drug-resistant epilepsy. Amygdala electrodes were implanted because semiological, EEG or MRI, results required clarification of whether the amygdala was part of the ictal onset zone. On average, the monitoring program lasted about one week. Written informed consent was given before participation. The Ethics Committee of Bielefeld University approved the study protocol (no. EUB 2016–115). Data were analyzed from 3 amygdala contacts (two bipolar channels) per patient, 10 amygdalae in total (6 right, 4 left). Some patients presented with bilateral amygdala implantations. However, unfortunately, only the signal from one hemisphere was clean enough for analysis. Given the scarcity of pre-surgical implantations, which are exclusively clinically motivated, we did not perform statistical tests to determine sample sizes.

However, our sample is similar to those reported in previous publications (Guex et al., 2020; Méndez-Bértolo et al., 2016). Approximate contact locations are detailed in Table 1 and Figure 1.

**Table 1**

Clinical details of the patients.

| Patient | Age, sex | Surgery – histopathology – Engel outcome (follow-up duration) | Amygdala pathology (MRI) | Amygdala Contacts (MNI Coordinates) | % trials free of interictal discharge |
|---|---|---|---|---|---|
| 1 | 29, F | R temp-lat topectomy, amygdala and hippocampus spared – mMCD II – IIIA (2 y) | None | CAR1 (22/-2/-16) CAR2 (26/0/-18) CAR3 (30/0/-18) | 97.79 |
| 2 | 22, M | Extended lesionectomy R temp-post – ganglioglioma – IA (2 y) | None | CAR1 (24/2/-18) CAR2 (28/2/-16) CAR3 (32/2/-16) | 92.23 |
| 3 | 21, M | Extended lesionectomy L temp-ant, amygdala and hippocampus spared – gliosis – IA (2 y) | None | CAL1 (-18/0/-24) CAL2 (-22/0/-24) CAL3 (-26/0/-26) | 54.55 |
| 4 | 33, M | R ant-med temp resection – hippocampal sclerosis, chronic encephalitis (anti-Ma2) – IIB (2 y) | Slight R FLAIR signal increase | CAL1 (-22/-4/-18) CAL2 (-26/-4/-18) CAL3 (-30/-4/-20) | 88.89 |
| 5 | 63, F | R ant-med temp. resection – hippocampal sclerosis – IIB (2 y) | None | CAR1 (26/-4/-20) CAR2 (30/-2/-22) CAR3 (34/0/-22) | 92.23 |
| 6 | 19, F | Extended lesionectomy L temp-lat-basal – ganglioglioma – IA | None | CAL1 (-18/-6/-24) CAL2 (-22/-4/-26) CAL3 (-26/-4/-26) | 85.56 |
| 7 | 43, M | None (bilateral sclerotic hippocampy, both generated seizures) | None | CAR1 (18/-6/-16) CAR2 (24/-4/-16) CAR3 (28/-4/-18) | 99.16 |
| 8 | 60, M | R ant-med temp resection – gliosis – IIIA (6 mo) | Slight R FLAIR signal increase | CAL1 (-16/-6/-20) CAL2 (-20/-4/-20) CAL3 (-24/-4/-20) | 99.16 |
| 9 | 21, M | Extended lesionectomy R fr – FCD IIA | None | CAR1 (18/-6/-16) CAR2 (22/-6/-16) CAR3 (26/-6/-18) | 96.39 |
| 10 | 56, M | Extended lesionectomy in the form of a R anteromedial temporal lobe resection - meningoencephalocele | None | CAR1 (14/-6/-20) CAR2 (20/-4/-20) CAR3 (26/-4/-22) | 80.56 |

*Notes*: Only trials that were free of interictal discharge in the amygdala were used for analysis. Abbreviations: ant = anterior; CAR = corpus amygdaloideum right; CAL = corpus amygdaloideum left; Engel outcome = outcome with respect to epileptic seizures (Engel, 1993). F = female; FCD = focal cortical dysplasia (Najm et al., 2022); FLAIR = Fluid-attenuated inversion recovery; fr = frontal; L = left; lat = lateral; M = male; med = medial; mMCD II = mild malformation of cortical development type II (Palmini et al., 2004); mo = months; R = right; temp = temporal; y = years.

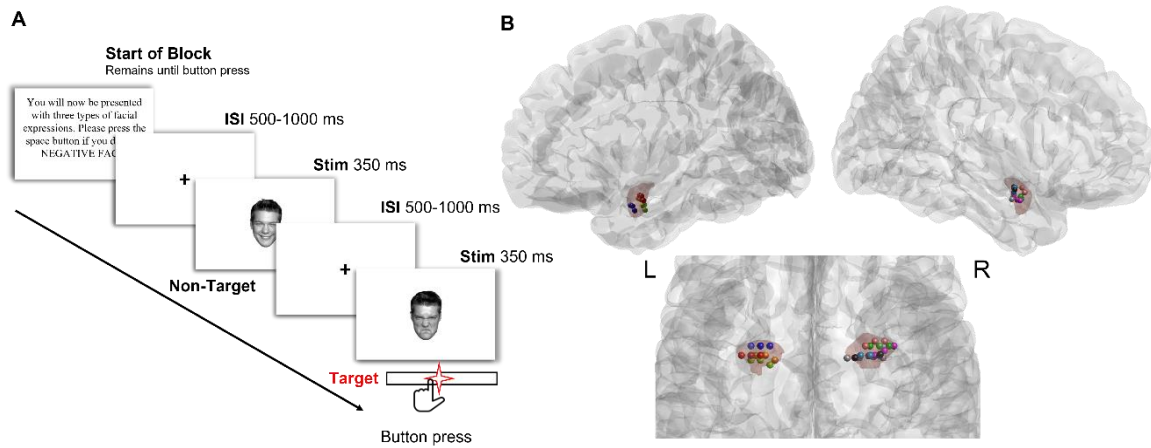

*Figure 1.* Experimental procedure and contact visualization. **A** Illustration of the experimental runs. After reading the instructions that indicated the target valence, participants viewed a serial presentation of three different facial expressions separated by a fixation cross in the middle of the screen. They were instructed to press the space bar when a target expression was detected. **B** Normalized contact positions from the analyzed stereo-encephalography strips. Images show approximated normalized depth contact positions from each patient (color coded) overlaid onto a normalized anatomical brain template, constructed with FreeSurfer software (version 7.3.2-20220804, Fischl, 2012). The amygdalae are highlighted in red. Abbreviations: L = left, ISI = Interstimulus-Interval, R = right, Stim = Stimulus.

*Healthy controls (HC).* The initial sample consisted of 25 participants. After excluding data from 6 participants due to large movement and perspiration artifacts, an analysis sample of 19 participants remained (age: *M* = 27.39 years, range = 19-52, 57.89 % female). All participants were right-handed, and all were native German speakers. No participant had a reported history of acute or past neurological or psychiatric disorders. The Beck Depression Inventory (BDI; *M* (*SD*) = 4.58 (3.17); Beck et al., 1996) and the State and Trait Anxiety Inventory (STAI; X1: *M* (*SD*) = 33.47 (10.46), X2: *M* (*SD*) = 33.84 (12.80); Spielberger, 1983) were used for screening of depressive symptoms and anxiety disorders and revealed no clinically relevant scores. All participants gave written informed consent before participation and received course credit.

Attention tunes gamma-band responses to faces

**Stimuli**

Forty identities, 50 % female, Western Caucasian faces were derived from the pool of "young" faces from the "FACES" database of the Max Planck Society for the Advancement of Science with previous permission (Ebner et al., 2010). Each identity was selected to show all three targeted expressions (angry, neutral, and happy). Only heads, in greyscale (see Figure 1), were placed centrally on a white background and scaled to a uniform width (9.60° vertical visual angle (centered)), naturally varying in height depending on head size. In each block, participants were instructed to actively monitor for one specific expression as the target condition, and to press the space bar when the target emotion was presented, while disregarding non-target expressions.

**Procedure**

Stimulus presentation was controlled using Presentation® Software (Version 18.1, Neurobehavioral Systems Inc., Berkeley, CA, http://www.neurobs.com) and carried out on a 15.6" laptop (DELL Latitude E5540, 1920 x 1080 pixels; DELL, Round Rock, TX, USA) for the patients. The control sample sat in front of a desk-mounted 14.1" Laptop (DELL Latitude D630, 1440 x 900 pixels; DELL, Round Rock, TX, USA) with an approximate screen distance of 50 cm in a laboratory at Bielefeld University. For the iEEG data, the laptop computer was positioned in front of the patients on their laps with a screen distance of ~50 cm while they sat on their hospital beds.

Participants were instructed to read the instructions on the screen carefully and avoid distractions during the presentation. Any further questions were addressed before starting the experiment. Trials were presented in three blocks, each instructing either negative, neutral, or positive faces as the target. Blocks were divided by breaks which the participants could pace individually. Each block started with instructions in black font (Times) on a white background in the middle of the screen. One block consisted of 120 trials, with 40 target stimuli and 80 non-targets that were randomly presented, adding up to 360 trials in total. Block order was counterbalanced across patients. Figure 1 shows an exemplary trial sequence.

Attention tunes gamma-band responses to facesEach trial consisted of a black fixation cross on a white screen with a variable interstimulus interval of 500 – 1000 ms to reduce stimulus onset predictability. The subsequent stimulus presentation was 350 ms. Participants were instructed to respond by pressing the space bar when a facial expression of the given target valence was detected. The following fixation cross indicated the beginning of a new trial. Figure 1 illustrates the trial sequence. Stimuli were repeated in each of the three blocks. Stimulus presentation was randomized across blocks. The experiment took about 15-20 minutes.

**Intracranial data acquisition and contact localization**

Patients were implanted with 5-9 stereotactic depth electrodes that contained 8-14 contacts each. An average of ~64 contacts in total were recorded. Electrodes were provided by Ad-Tech (Racine, WI, USA). The signal was continuously recorded using a Nihon-Kohden system (Tokyo, Japan) or a Compumedics Neuroscan system (Singen, Germany) with a sampling rate of 1000-5000 Hz, depending on the patient (online bandpass filter 0.1 – 200 Hz). The original recording was carried out with a bipolar montage of adjacent contacts building a recording channel. Additional channels for an electrocardiogram (ECG) and a synchronous recording of trigger pulses were also recorded.

The analyzed amygdala contacts were selected based on thorough visual inspection of the post-implantation MRI scans. Normalized coordinates of respective contacts were additionally used for localization validation by using independent meta-analytical data obtained from NeuroSynth (http://neurosynth.org; Yarkoni et al., 2011). Coordinates were obtained by performing a co-registration of preoperative MRI scans and post-operative CT scans implemented in SPM8 (Friston et al., 2007) and executed with the FieldTrip toolbox (version 20200121; Oostenveld et al., 2011) running on MATLAB (version 2019b, The MathWorks Inc.), as described in Stolk et al. (2018). The post-operative CT was transformed into the native space of the pre-operative MRI via affine normalization. Both images were then digitally fused and overlaid. Contacts, as visible in high-contrast post-implantation CT images, were individually and manually marked in the fused image. The contact locations were normalized to the MNI space, so that voxels in which contacts were localized could be used to determine the approximate MNI coordinates of the contacts. Four patients (patients 3, 4, 8, and 9) had insufficient imaging data for this method of contact localization. Here, we normalized the post-operative MRI scan with SPM8 using FieldTrip scripts and extracted the MNI voxel coordinates that contained the contacts of interest in 3D Slicer (version 5.0.3; Kikinis et al., 2014). Three amygdala contacts per patient were used for analysis (see Table 1).

**iEEG data processing**

Data processing was carried out using FieldTrip software running on MATLAB 2019b. For offline processing of the time-frequency data, the signal was down-sampled to 500 Hz. A bipolar montage was applied for the analysis, each channel consisting of two adjacent contacts (i.e., CAR1-CAR2; CAR2-CAR3, or CAL1, etc., for left-hemispheric contacts). The signal was high-pass filtered with a second-order zero-phase Butterworth filter at 0.1 Hz, a second-order zero-phase low-pass Butterworth filter of 200 Hz was also applied. A notch filter additionally removed power-line noise from 49-51 Hz (for a similar approach, see Guex et al., 2020). The single trial epochs were determined using a 700 ms pre-stimulus and a 1000 ms post-stimulus interval. Epochs containing large artifacts from interictal epileptic discharges in the time-domain (epileptic bursts, low-frequency high amplitude oscillations, high-frequency noise) were excluded from analysis after careful trial-by-trial visual inspection.

We calculated the time-frequency power spectrum from 35 to 90 Hz (low gamma, LG) in steps of 2.5 Hz. Since emotion effects in iEEG studies were previously found in higher gamma frequencies (Boucher et al., 2015; 2022; Guex et al., 2020), we additionally calculated high gamma (HG) power from 90-150 Hz using the same parameters as for LG. We used Slepian tapers with a fixed window length of 400 ms and applied spectral smoothing through multi-tapering (7 tapers) of 20 Hz for all frequencies (for a similar approach, see Moratti et al., 2014). This way, we obtained an equal number of time-frequency bins across the time-frequency spectrum for statistical analysis (Gross et al., 2013). A baseline correction for the frequency calculations was applied by expressing the gamma-power as a relative change to the power in the baseline of 700 to 200 ms pre-stimulus interval. Epochs containing large artifacts from interictal activity in the time-frequency domain (epileptic bursts, low-frequency high amplitude oscillations, high-frequency noise) that were not previously detected in the time domain were excluded from analysis after careful trial-by-trial visual inspection. Finally, data were averaged across trials and channels for each condition and each patient. On average, 5.40 % of trials were rejected per condition. The rejection rate did not differ across conditions ($F_{(8,64)}$ = 0.492, $p$ = .621, $\eta_p^2$ = .051). The relative number of included trials can be found in Table 1.

**Scalp EEG data acquisition and processing**

Scalp EEG was continuously recorded from 32 active BioSemi Ag-AgCl electrodes using the ActiView software (http://www.biosemi.com). The horizontal electrooculogram (EOG) was measured with two electrodes placed on the outer canthi of the left and right eye. Two further electrodes for the vertical EOG were placed over and under the left eye. The recording was referenced to the Cz electrode, with the sampling rate set to 1024 Hz (online bandpass filter 0.1 – 200 Hz). Impedances were kept below 25 Ω, in line with system recommendations. Pre-processing of offline data was performed using the FieldTrip toolbox running in MATLAB 2019b. Offline data were re-referenced to the common average, and data were down-sampled to 500 Hz. In line with recent research about scalp GBA (e.g., Yang et al., 2020), a high-pass Butterworth second-order zero-phase filter of 0.1 Hz and a low-pass second-order zero-phase Butterworth filter of 100 Hz were applied. A notch filter additionally removed power-line noise from 49-51 Hz. Filtered data were cut into segments of -700 ms to 1000 ms relative to stimulus onset. Large muscle or technical artifacts were removed manually after careful trial-by-trial visual inspection. Further artifact correction was performed using the independent component analysis (ICA) runICA algorithm (Makeig et al., 1995). On average, 7 components were rejected per participant. Data were averaged across trials per participant for down-weighing possible residual artifacts. No electrodes were interpolated.

Face perception was previously associated with cortical GBA increases from 55-71 Hz (Moratti et al., 2014), and EEG studies of face emotion effects in scalp GBA reported effects from 35-80 Hz (e.g., Müller et al., 2000; Yang et al., 2020). Therefore, we calculated the time-frequency power spectrum from 35 to 90 Hz in steps of 2.5 Hz. Higher frequencies were not considered for the scalp data, because of the filtering characteristics of the skull (Srinivasan et al., 1996). As in the iEEG time-frequency calculation,

Attention tunes gamma-band responses to faces we used Slepian tapers with a fixed window length of 400 ms and applied spectral smoothing through multi-tapering (7 tapers) of 20 Hz for all frequencies. A baseline correction for the frequency calculations was applied by expressing the gamma-power as a relative change to the power in the baseline of 700 to 200 ms pre-stimulus interval. Epochs containing large artifacts in the time-frequency domain (technical noise, muscle artifacts) that were not previously detected in the time domain were excluded from analysis by careful trial-by-trial by visual inspection. Finally, data were averaged across trials for each condition and each subject. On average, 4.62 % of trials were rejected per condition. The rejection rate did not differ across conditions ($F_{(8, 144)}$ = 0.889, $p$ = .526, $\eta_p^2$ = .04).

**Data analysis**

**Behavioral data**

Data were analyzed using R Software (R Core Team, 2013). The detection accuracy was calculated by computing the difference between hits and false alarms relative to the total number of targets or non-targets, respectively. Reaction times were calculated by measuring the time between stimulus onset and space button press. Data were then averaged for each target valence and each subject. With mixed-design ANOVAs, we tested for main effects and an interaction of the factors *group* (HC versus patients) and *target condition* (attend to negative versus attend to neutral versus attend to positive) in the detection accuracy and reaction times. In case of significant effects, we compared the data post-hoc between the groups with independent t-tests. The three target conditions were compared with dependent t-tests. Assumptions of normality and sphericity were tested beforehand. Greenhouse-Geisser corrections of degrees of freedom were applied in case of a violation of the sphericity assumption. For readability, uncorrected degrees of freedom but corrected p-values are reported. The normality assumption of differences for paired t-tests was tested beforehand. Holm-corrections were applied to account for the heightened probability of false positives (Holm, 1979). To ensure a more unbiased interpretation of the data we additionally report partial eta-squared ($\eta_p^2$) for all parametric F-tests (Cohen, 1973). Following Cohen (1988), effects are interpreted as small when $\eta_p^2$ > .01, medium when $\eta_p^2 > .07$ and large when $\eta_p^2 > .13$. For parametric t-tests, Cohen's d was reported as effect size with d = .20 being a small, d = .50 medium, and d = .80 a large effect (Cohen, 1992).

**iEEG analysis**

Statistical data analysis of the iEEG data was performed using FieldTrip software running on MATLAB 2019b. For all iEEG analyses, we searched for effects in the average signal of two bipolar amygdala channels per patient.

*Effects for target runs and targets.* First, to test for systematic differences between the target conditions, we analyzed the gamma-band activity in LG (35-90 Hz) and HG (90-150 Hz) separately, testing for differences between the three target conditions (factor *target condition* (attend to negative versus attend to neutral versus attend to positive)), pooled across facial expressions. Furthermore, we were interested in effects of the factor *target* (angry versus neutral versus happy) that would indicate differential target enhancements based on valence. For this, we computed pointwise dependent cluster-based permutation F-tests. We conducted parametric dependent two-tailed t-tests to compare GBA in a given time-frequency cluster between the target conditions and the targets when they differed significantly.

*Within-condition analysis.* Effects of the factor *expression* (angry versus neutral versus happy) within the target conditions were furthermore tested in LG and HG with pointwise dependent cluster-based permutation F-tests. We conducted parametric dependent two-tailed t-tests to compare GBA in a given time-frequency cluster between the target conditions and the targets when they differed significantly. Additionally, these analyses were carried out on a single-patient level with independent F-Tests (see supplementary material), treating the single trials as separate and independent cases (Maris & Oostenveld, 2007).

Non-parametric cluster-based permutation F-tests were performed using the Montecarlo method and a maximum summation approach (Maris & Oostenveld, 2007). Data at each time-point was randomly shuffled between conditions, and after each permutation step, an F-value was calculated for each time-point from 0-800 ms in steps of 10 ms. Note that for some frequencies, the very edge of the spatio-temporal reconstruction (795-800 ms) was not calculated and, therefore, not included in the statistical analysis. No degrees of freedom are given for non-parametric measures. Significant clusters were defined by temporal and frequency adjacency with a cluster threshold of alpha = .05, correcting for multiple comparisons on the cluster-level. Permutation steps were repeated 1000 times. For each cluster, the F-values were summed and the greatest sum among all clusters was entered into the permutation distribution. Maximum F-values with $p \leq .05$ were considered as statistically significant.

**Scalp EEG Analysis**

For the scalp EEG, we conducted the same analyses as for the iEEG (see above). However, all differences between conditions were only tested in a frequency range from 35-90 Hz (LG) but not 90-150 Hz (HG). To investigate the spatial distribution of GBA effects on the scalp, we did not average the signal across channels for each subject as we did for the amygdala channels.

## RESULTS

### Behavioral data

Table 2 and Figure 2 illustrate the behavioral data.

**Table 2**

Mean reaction times and hit rates for each experimental block.

|  |  | Target Condition | | |
| --- | --- | --- | --- | --- |
|  |  | negative | neutral | positive |
| Patients | RT (*M* (*SD*)) | 541.99 (88.92) | 616.97 (106.47) | 533.71 (106.04) |
|  | Detection Accuracy (*M* (*SD*)) | 82.50 (17.13) | 80.12 (17.62) | 89.50 (12.16) |
| Healthy Controls | RT (*M* (*SD*)) | 537.03 (72.67) | 570.71 (61.62) | 488.31 (72.72) |
|  | Detection Accuracy (*M* (*SD*)) | 92.5 (4.70) | 92.94 (5.36) | 97.57 (3.49) |

*Notes.* RT = Reaction Times, *M* = Mean, *SD* = standard deviation. Detection accuracies are expressed as the number of hits -

Attention tunes gamma-band responses to faces false alarms relative to the absolute number of targets or non-targets, respectively.

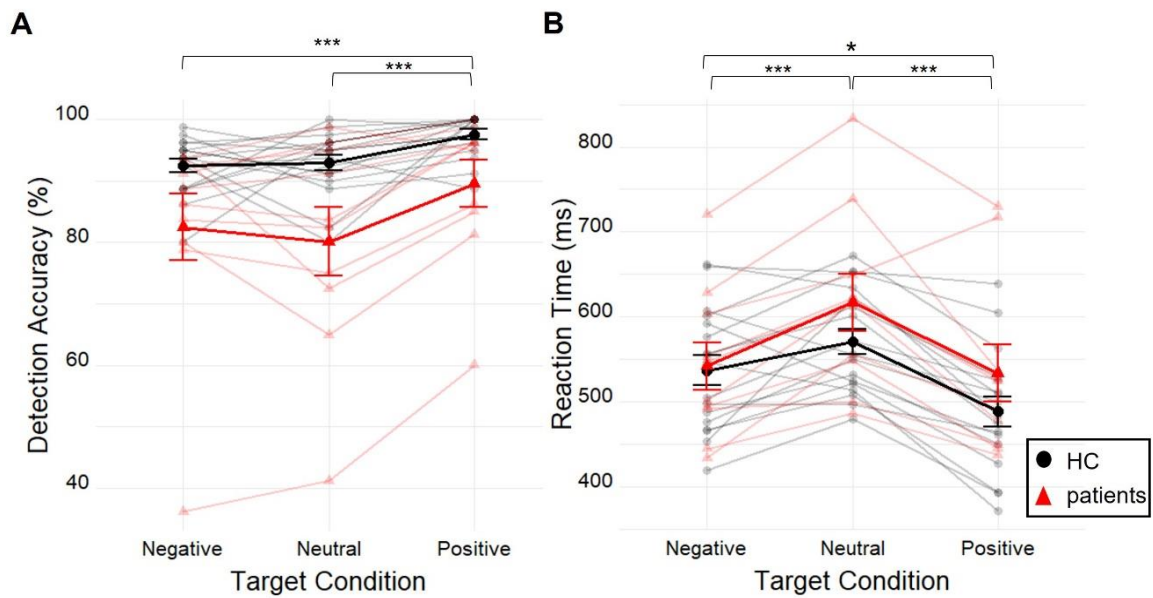

*Figure 2.* Behavioral data of the patient (triangles) and the HC group (circles). (**A**) Mean detection accuracy within the three target conditions is expressed as hits - false alarms, relative to the absolute number of targets or non-targets, respectively. (**B**) Mean reaction times (in ms) within the three target conditions. Opaque lines show the mean behavioral performance per group, and transparent lines show the single subjects. Error bars indicate the standard error of the mean (SE). Brackets indicate significant (*$p$ ≤ .05, ***$p$ ≤ .001) within- or between-group differences.

The mixed-model ANOVA revealed a significant main effect of the factor *group* ($F_{(1,25)}$ = 7.75, $p$ = .001, $\eta_p^2$ = .236) and a significant main effect of the factor *target condition* ($F_{(2,50)}$ = 12.719, $p$ < .001, $\eta_p^2$ = .33) for the detection accuracy. Post-hoc tests specified that the HC group detected the targets more accurately than the patient sample ($t_{(32.5)}$ = 3.47, $p_{corr}$ = .001, $d$ = .991), although overall performance was very high (> 80 %) in both groups. Furthermore, happy faces were detected more accurately than neutral ($t_{(26)}$ = 4.98, $p_{corr}$ < .001, $d$ = .583) and angry faces ($t_{(26)}$ = 4.28, $p_{corr}$ < .001, $d$ = .559). The detection accuracy for the angry and neutral faces did not differ significantly ($p_{corr}$ = .706).

For the reaction times, the mixed-model ANOVA revealed a significant main effect of the factor *target condition* ($F_{(2,50)}$ = 28.498, $p$ < .001, $\eta_p^2$ = .532). Reaction times in the *attend to positive* condition were faster compared to the *attend to neutral* condition($t_{(26)}$ = 8.32, $p_{corr}$ < .001, $d$ = .973), and so were

Attention tunes gamma-band responses to faces reaction times in the *attend to negative* condition ($t_{(26)}$ = 4.62, $p_{corr}$ < .001, $d$ = .612). Reaction times in the *attend to positive* condition were additionally faster than those in the *attend to negative* condition ($t_{(25)}$ = 2.59, $p_{corr}$ = .015, $d$ = .408).

**Patient sample: intracranial data**

**Overall differences in GBA between individual target runs and emotional expressions**

Shown in Figure 3 are grand-average time-frequency plots that illustrate the grand-average power values from each condition.

The cluster-based permutation tests revealed no effect of the factor *target condition* ($ps_{corr}$ > .611), indicating no general differences in either LG or HG between the attentional conditions. Likewise, cluster-based permutation tests revealed no significant differences in LG or HG between the targets (all $ps_{corr}$ > .118).

**Gamma modulation within the attentional conditions**

Plots and statistics illustrating the analyses can be found in Figure 3.

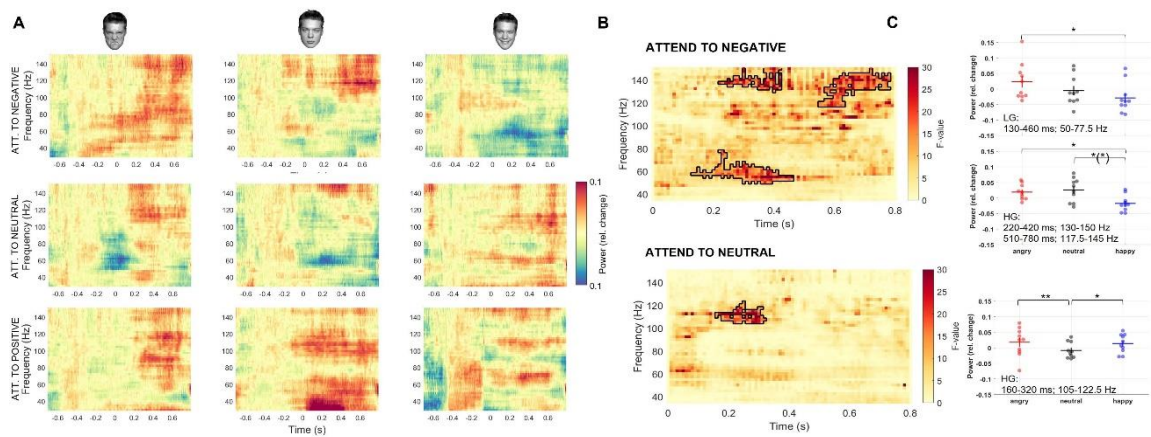

*Figure 3.* Significant effects in the *attend to negative* and *attend to neutral* conditions in the amygdala. (**A**) Grand-average time-frequency plots of gamma-band power in the amygdala from all conditions. (**B**) Time-frequency plots of F-values for the within-condition analyses. Significant clusters that survived the correction for multiple comparisons are outlined in black. Note that the early HG cluster during attention to negative faces only reached $p$ = .053. (**C**) Swarm plots of single-subject mean power values

(see also Supplementary Table 1) within time-windows of significant effects. Horizontal lines: group means. Error bars: standard error of the mean. Brackets indicate significant differences as revealed by post hoc tests (*$p \leq .05$, **$p \leq .01$). Due to a similar pattern of activation, the two HG clusters in attend to negative were concatenated. Thus, the brackets indicate results from two post-hoc tests. Abbreviations: Att. To = Attend to.

*Attend to negative.* Cluster-based permutation tests revealed a significant effect of the factor *expression* in LG (summed cluster *F*-value = 2260.90, $p_{corr}$ = .042) from 130-460 ms in the frequency range of 50-77.5 Hz. An effect in HG from 220-420 ms in the frequency range of 130-150 Hz closely failed to reach significance (summed cluster *F*-value = 1700.40, $p_{corr}$ = .053) but another significant effect arose from 510-780 ms in the frequency range of 117.5-150 Hz (summed cluster *F*-value = 2723.30, $p_{corr}$ = .034). In the low GBA cluster, GBA was significantly higher for angry compared to happy faces ($t_{(9)}$ = 3.337, $p_{corr}$ = .024, $d$ = .781). GBA did not differ significantly between angry and neutral faces, nor between neutral and happy faces (all $ps_{corr}$ = .160). In the both high GBA clusters, GBA was significantly higher for angry and neutral faces compared to happy ones (early: angry > happy: $t_{(9)}$ = 3.304, $p_{corr}$ = .018, $d$ = .755; neutral > happy: $t_{(9)}$ = 4.487, $p_{corr}$ = .003, $d$ = .957; late: angry > happy: $t_{(9)}$ = 3.681, $p_{corr}$ = .015, $d$ = .838; neutral > happy: $t_{(9)}$ = 2.905, $p_{corr}$ = .034, $d$ = .710). GBA did not differ significantly between angry and neutral faces (all $ps_{corr}$ = .200).

*Attend to neutral*. Cluster-based permutation tests revealed no significant effects in LG (all $ps_{corr}$ > .410). In HG, the analysis revealed a significant effect of the factor *expression* from 160-320 ms in a frequency range of 105-122.5 Hz (summed cluster *F*-value = 1563.60, $p_{corr}$ = .049). Here, GBA was significantly increased for angry and happy compared to neutral faces (angry > neutral: $t_{(9)}$ = 4.305, $p_{corr}$ = .006, $d$ = .927; happy > neutral: $t_{(9)}$ = 3.229, $p_{corr}$ = .018, $d$ = .759). GBA in response to angry and happy faces did not differ significantly ($p_{corr}$ = .525).

*Attend to positive*. No significant effects of the factor *expression* arose within the *attend to positive* condition in LG and HG (all $ps_{corr}$ > .398).

Attention tunes gamma-band responses to faces

Individual power values that constitute the significant effects can be found in Supplementary Table 1.

**HC group: Scalp data**

**Overall differences in GBA between individual target runs and emotional expressions**

Cluster-based permutation tests revealed no significant effect of the factor *target condition,* nor did they show significant overall differences in gamma-band power between the targets (all $ps_{corr}$ > .167).

**Gamma modulation within the attentional conditions**

Time-frequency plots and statistics illustrating the analyses can be found in Figure 4.

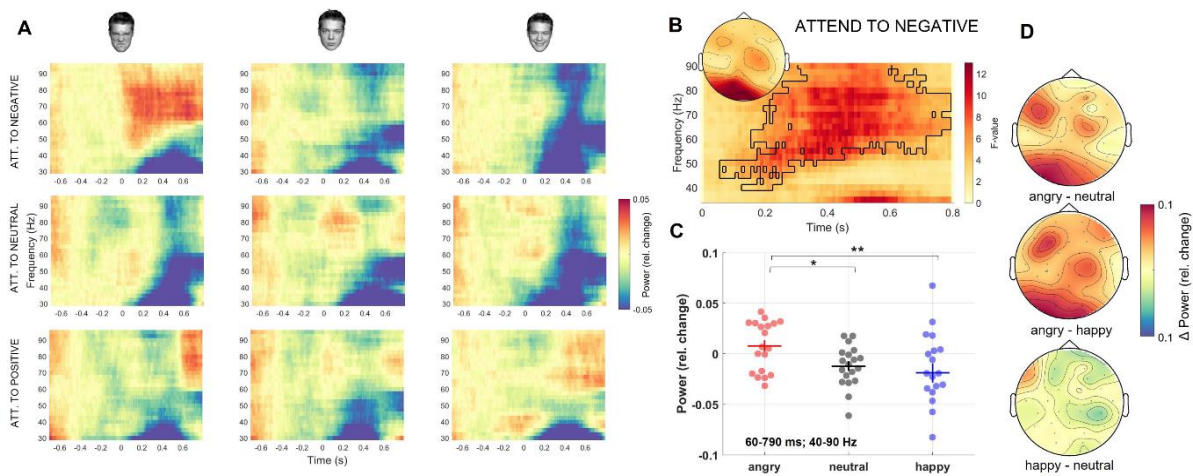

*Figure 4.* Significant effects in the *attend to negative* condition on the scalp. (**A**) Grand-average time-frequency plots of scalp gamma-band power from all conditions, averaged across all scalp channels. (**B**) Time-frequency and channel-frequency plots depicting the F-values within the *attend to negative* condition. Significant clusters that survived the correction for multiple comparisons are outlined in black. (**C**) Single-subject mean power values within time-windows of significant effects, averaged across channels, time-points, and frequency-bins. Horizontal lines: group means. Vertical lines: standard error of the mean. Brackets indicate (marginally) significant differences as revealed by the post hoc tests (**$p ≤ 01$, *$p ≤ .05$). (**D**) Grand-average condition differences, averaged over time-points and frequency-bins of the significant effect.

*Attend to negative.* Cluster-based permutation tests revealed a significant effect of the factor *expression* from 60-790 ms in a frequency range of 40-90 Hz (summed cluster *F*-value = 68,930, $p_{corr}$ = .002). It was most concentrated at posterior channels but covered a large area of the scalp (see also Figure 4). T-tests revealed that gamma-power was significantly higher for angry than happy faces ($t_{(18)}$ = 3.386, $p_{corr}$ = .009, $d$ = .635) and higher for angry than neutral ($t_{(18)}$ = 2.416, $p_{corr}$ = .050, $d$ = .468) faces. Gamma-band power in response to neutral and happy faces did not differ significantly ($p_{corr}$ = .408).

*Attend to neutral.* No significant effects of the factor *expression* arose within the *attend to neutral* condition (all $ps_{corr}$ > .539).

*Attend to positive.* No significant effects of the factor *expression* arose within the *attend to positive* condition (all $ps_{corr}$ > .386).

## DISCUSSION

Our aim was to extend previous findings on interactions of emotional and directed attention in face processing in the amygdala and on the scalp, specifically regarding gamma-band oscillatory responses. Many studies focused on fear processing and its interactions with top-down attention, whereas we aimed to go beyond fear processing, including angry and happy faces. Our previous single-case iERP study suggests that attention affects differentiation of angry, neutral, and happy expressions earlier in the amygdala than on the scalp (Weidner et al., 2022), implying early attentional guidance of amygdala emotion differentiation. However, Guex et al. (2020) demonstrated a divergence between amygdala iERPs and GBA regarding top-down attentional modulation. Similarly, Engell and McCarthy (2010) assume a functional divergence of cortical ERPs and GBA. Hence, we focused on GBA, thereby extending previous ERP research, and closing a gap between findings regarding GBA on the scalp and in the amygdala.

Present behavioral data indicate a recognition advantage for emotional over neutral faces. This is in line with a large body of other studies (for review see Xu et al., 2021). According to Hodsoll et al. (2011) and Schwabe et al. (2011), this advantage might be caused by an automatic attentional shift in favor of emotionally arousing faces. The additional benefit for happy compared to angry faces was likely

Attention tunes gamma-band responses to faces driven by their high emotional distinctiveness (Becker et al., 2011; Bucher & Voss, 2019; Calvo & Beltrán, 2013).

When negative expressions were task-relevant, both scalp (~60 ms) and amygdala (~130 ms) low GBA were selective to angry faces, and high GBA in the amygdala was persistently increased for angry and neutral compared to happy faces at 220-420 ms and 510-780 ms. When participants monitored for neutral expressions, specifically amygdala GBA increased in response to emotional faces as compared to neutral faces at 160 ms. Assuming that GBA indicates attentional guidance (Jensen et al., 2012; Jensen et al., 2014), this might reflect automatic attentional shifts towards affective relevance, which we will elaborate on further below. Attention to positive faces did not differentiate expressions on the group-level. Thus, behavioral and GBA data seem to reflect functionally different mechanisms of emotional face detection. Single-subject statistics of iEEG data demonstrated interindividual variability in the context-specificity of emotion-driven amygdala responses.

*Attention to threat suppresses the detection of task-irrelevant emotion*

The explicit instruction to monitor for negative faces resulted in stronger low GBA synchronization to angry faces, both on the scalp around 60 ms and in the amygdala later at 130 ms. In amygdala high GBA, angry and neutral faces were also consistently differentiated from happy faces from 220 ms on. As no effects were found in the *attend to positive* condition, target effects for negative faces did not generalize to attention to emotion in general. This extends work from Müsch and colleagues (2014), who did not report differentiations between fearful and neutral facial expressions in amygdala GBA when happy expressions were task-relevant. They argue for a potential overshadowing of emotion effects by "face attention" in general. We show that attention to faces does not universally enhance amygdala GBA but interacts with affective relevance of the attended to face. Our data suggest a mechanism of cortical and subcortical input integration carried out through neural synchronization in the gamma range (Balconi & Lucchiari, 2008; Moratti et al., 2014; Luo et al., 2007, Luo et al., 2009) that is specifically tuned by attention to negative faces. In accordance, Herz et al. (2020) theorize that negative affect narrows cerebral processing in favor of negative stimuli, while positive affect broadens it, which could have driven the less selective face processing during attention to positive faces.

Although definitive assumptions about the onset of an effect are difficult due to constraints in interpreting the timing in cluster-based analyses (Maris & Oostenveld, 2007), and the data originates from two different samples, our results show that differentiation of angry targets on the scalp preceded respective differentiation in the amygdala. This does not support the hypothesis that visual information about affective relevance travels to the amygdala through a subcortical shortcut (Garvert et al., 2014; LeDoux, 2007; McFadyen et al., 2017) but rather through re-entrant connections from the visual cortex (Vuilleumier, 2005). In line with the re-entrant idea (Vuilleumier, 2005), the effects were predominantly present in earlier stages of perceptual processing. Single-cell recordings in monkeys suggest that top-down attention activates feedback signals between lower and higher parts of the ventral visual system (Debes & Dragoi, 2023), and such perceptual sensitivity might be particularly increased during attention to potential threat (Voogd et al., 2022). Also, prefrontal signals, which guide attention (Ishii et al., 1999; Small et al., 2003), might have selectively contributed to a widespread GBA synchronization, including the visual cortex (Desimone & Duncan, 1995) and the amygdala (for review see Kaldewaij et al., 2016), during threat vigilance. This hints toward an evolutionarily specialized mechanism that facilitates processing during monitoring for negative social information (Pratto & John, 1991). Attention to potential threat may even suppress some pre-attentive effects of emotional processing in the amygdala, as it diminished the differentiation of happy faces which were differentially processed during attention to neutral faces.

Interestingly, previous research reported differentiations of negative and neutral faces in amygdala high GBA (60-150 Hz) (Guex et al., 2020; Guex et al., 2022; Zheng et al., 2017). By contrast, at least in the *attend to negative* condition, we found little differentiation of angry and neutral faces either in low or high amygdala GBA. Of note, angry faces were also distinguished more clearly from happy faces than from neutral ones in the behavioral data, suggesting classification ambiguities between angry and neutral faces. Neutral faces could have made an initial impression as negative (Albohn et al., 2019; Lee et al., 2008), as we also discussed in our previous ERP study (Weidner et al., 2022). Therefore, the amygdala might be recruited during evaluation of ambiguous social information (Wang et al., 2017). Inclusion of happy faces in the paradigm might have enhanced ambiguities between angry and neutral ones, as positive faces have a more distinct and recognizable expression (Becker et al., 2011).

*A neutral focus enhances the detection of emotion in the amygdala*

Rapid, attention-independent emotion detection in the amygdala was observed solely during a neutral focus. Similarly, Guex et al. (2020) reported an early-onset (80 ms) GBA increase to fearful relative to neutral faces when neutral faces were task-relevant. Processing interferences by task-irrelevant emotional faces were repeatedly demonstrated in the amygdala (for review see Dolan & Vuilleumier, 2003). But, while most studies show this to be fear-specific, our results demonstrate that a bias for task-irrelevant emotion is present for threatening and positive faces as well. Since the behavioral performance was also poorest for neutral targets and potentially disturbed by emotional distractors (Fox et al., 2002; Hodsoll et al., 2011) the amygdala might be part of a distributed network that contributes to an involuntary attentional bias for emotion, ensuring detection of unattended emotion (Ciesielski et al., 2010; Kennedy & Most, 2015; Mitchell et al., 2008; Most et al., 2007; Schwabe et al., 2011; for review see McHugo et al., 2013). It most likely does not drive this bias on its own, as some show a preserved automatic attentional capture by emotional distractors in patients with uni- and bilateral amygdala damage (Piech et al., 2011; Tsuchiya et al., 2009).

Our data imply distinct neural pathways of emotion processing on the cortex and in the amygdala during a neutral focus, as the differentiation of task-irrelevant emotion was not found on the scalp. Interestingly, this contrasts with scalp ERP data from this sample that show such differentiation (Weidner et al., 2022), which further corroborate that distinct neural mechanisms generate ERPs and GBA (Engell & McCarthy, 2010). Because of the dissociation of amygdala and scalp GBA, and amygdala GBA being unaffected by top-down attention, bottom-up processes seem to have driven facial emotion recognition. These do not entail cortical feedback signals and might be contributed to by subcortical, magnocellular connections to the amygdala, although these connections were primarily assumed to be relevant during fear processing (LeDoux, 2007; Méndez-Bértolo et al., 2016; Vuilleumier et al., 2003). However, Wyatte et al. (2014) demonstrated very fast (80-100 ms) feedback between lower and higher parts of the ventral visual stream that might also intrinsically encode stimulus valence (Miskovic & Anderson, 2018) but these feedback connections seem to be less active for unattended stimuli (Debes & Dragoi, 2023). Still, re-entrant connections from the visual system might be able to rapidly signal emotional relevance to the amygdala (Vuilleumier, 2005) which could have been overlooked due to methodological limitations regarding high gamma analyses in scalp EEG (Srinivasan et al., 1996).

Single-subject statistics (Supplementary Table 2, 3, 4, and 5) for the patients revealed interindividual variability regarding differentiations of expressions in GBA. Seven patients showed an emotion-specific GBA increase, varying in frequencies, time-windows, and the target condition in which these effects occurred. Although the sample is too small for definitive assumptions, these analyses reveal no clear functional distinctions between left and right amygdala in processing of emotion (e.g., Sergerie et al., 2008). However, we cannot rule out lateralization of effects within single patients, which, unfortunately, we cannot test for due to the lack of usable bilateral implantations.

**Limitations and Outlook**

Some limitations must be taken into consideration when interpreting our results. Generally, using a clinical sample can limit generalizability since the patients' disorders may influence external validity (e.g., disruption of cognitive functions, Dinkelacker et al., 2016; Nicolai et al., 2012). However, the present patients appeared cognitively intact, and through careful inspections of the data, we were able to exclude interfering epileptic activity from the signal. Additionally, compared to the healthy control group, patients for whom this information was available had elevated scores on the BDI that indicated presence of depressive symptoms in almost all patients. Comorbidities of depression and epilepsy are common (Korczyn et al., 2013) which the pre-surgical monitoring situation might have exacerbated. However, previous studies on the relationship of depressive symptoms with facial emotion recognition are contradictory (Bonora et al., 2011; Leppänen et al., 2004).

Attention tunes gamma-band responses to facesFurthermore, assumptions about possible influences of amygdala signals on cortical processing cannot be formally tested, as no recordings of cortical signals were carried out in the patient sample, and we derived our conclusions about the association of amygdala and scalp emotion responses from two different samples. Also, since we could not reliably test for effects in higher GBA on the scalp due to low pass filtering of the skull (Srinivasan et al., 1996), simultaneous intracranial recordings of visual cortices should be carried out to further investigate whether emotion-driven changes in amygdala GBA origin from cortical or subcortical areas. Additionally, we considered the amygdala as a whole, not differentiating the amygdala subnuclei that have specific functional characteristics (Bzdok et al., 2013; Zhang et al., 2023). This might be entangled with single-cell recordings in future research.

**Summary**

In sum, emotion-driven increases in GBA *can* occur automatically and independently of the attentional focus, but this seems to depend on the attended to expression. We refine knowledge from scalp (Balconi & Lucchiari, 2008; Luo et al., 2009) and amygdala (Guex et al., 2020) GBA studies regarding the attention-(in)dependence of emotional processing. These results provide possible explanations for ambiguous data regarding the attentional modulation of electrophysiological responses to faces. They demonstrate that emotion-sensitivity is highly context-dependent. The task-induced emotional focus represents one relevant context.

**SUPPLEMENTARY MATERIAL**

**Supplementary Table 1**

Individual gamma-power values from each patient in the time-frequency clusters of significant effects

| | Target Condition | | | | | | | | |
|---|---|---|---|---|---|---|---|---|---|
| | negative 130-460 ms; 50-77.5 Hz | | | negative 220-420 ms; 130-150 Hz 510-780 ms; 117.5-145 Hz | | | neutral 160-320 ms; 105-122.5 Hz | | |
| Patient | A | N | H | A | N | H | A | N | H |
| 1 (R) | 0.078 | 0.031 | 0.066 | 0.040 | 0.039 | -0.009 | 0.040 | 0.039 | -0.009 |
| 2 (R) | -0.012 | -0.038 | -0.053 | 0.022 | 0.050 | -0.048 | 0.022 | 0.050 | -0.048 |
| 3 (L) | 0.041 | -0.013 | -0.036 | -0.002 | 0.067 | -0.014 | -0.002 | 0.067 | -0.014 |
| 4 (L) | 0.025 | 0.062 | -0.040 | 0.012 | 0.021 | 0.021 | 0.012 | 0.021 | 0.021 |
| 5 (R) | -0.037 | -0.073 | -0.052 | -0.003 | -0.018 | -0.023 | -0.003 | -0.018 | -0.023 |
| 6 (L) | -0.022 | -0.014 | -0.078 | -0.014 | -0.018 | -0.047 | -0.014 | -0.018 | -0.047 |
| 7 (R) | -0.021 | -0.039 | -0.081 | 0.005 | 0.054 | -0.020 | 0.005 | 0.054 | -0.020 |
| 8 (R) | -0.024 | -0.035 | -0.039 | 0.020 | -0.029 | -0.023 | 0.020 | -0.029 | -0.023 |
| 9 (L) | 0.053 | 0.075 | 0.044 | 0.058 | 0.080 | 0.027 | 0.058 | 0.080 | 0.027 |

| | | | | | | | | |
|---|---|---|---|---|---|---|---|---|
| 10 (R) | 0.153 | -0.009 | -0.018 | 0.052 | 0.012 | -0.037 | 0.052 | 0.012 | -0.037 |

*Notes*. Values are depicted only for marginally significant or significant effects of the factor *expression* within the target conditions. Patients are sorted based on the hemisphere of their amygdala implantations. Abbreviations: A = angry, H = happy N = neutral.

**Single patients: Expressions within target conditions**

Supplementary Tables 2 and 3 illustrate the single-subject analyses in LG and HG that were carried out analogously to the analyses reported in the group statistics, except that single trials were treated as separate, independent cases. Statistical indices of the post-hoc tests can be found in Supplementary Tables 4 and 5. Overall, in LG, two patients showed effects in the *attend to negative* condition, two patients showed effects in the *attend to neutral* condition, and two patients in the *attend to positive* condition. In HG, two patients showed effects in the *attend to negative* condition and two in the *attend to neutral* condition. Individual effects were variable in onset and frequency-bands.

**Supplementary Table 2**

Significant single-subject comparison of the expressions within the target conditions (LG)

| Pat | TC | Freq (Hz) | Time (ms) | Cluster *F* | Cluster *p* | Post-hoc differences |
|---|---|---|---|---|---|---|
| 1 (R) | neutral | 52.50-67.50 | 8-486 | 6463.7 | .031 | angry > neutral |
| | | | | | | happy > neutral |
| 3 (L) | negative | 35-90 | 0-790 | 37675 | < .001 | angry > neutral |
| | | | | | | angry > happy |
| | neutral | 35-52.50 | 0-388 | 4645.4 | .036 | angry > happy |
| | | | | | | neutral > happy |
| 4 (L) | negative | 57.50-77.50 | 0-414 | 6105.2 | .013 | angry > happy |
| | | | | | | neutral > happy |
| | positive | 35-47.50 | 216-790 | 6043.8 | .016 | angry > neutral |
| | | | | | | happy > neutral |
| 6 (L) | positive | 35-47.50 | 122-782 | 3151.2 | .097 | happy > angry |
| | | | | | | happy > neutral |

*Notes*. Statistical data are depicted only for marginally significant or significant effects of the factor expression. Patients are sorted based on the hemisphere of their amygdala implantations. Abbreviations: L = left, Pat = patient, R = right, TC = target condition.

**Supplementary Table 3**

Significant single-subject comparison of the expressions within the target conditions (HG)

Attention tunes gamma-band responses to faces

| Pat | TC | Freq (Hz) | Time (ms) | Cluster F | Cluster p | Post-hoc differences |
|---|---|---|---|---|---|---|
| 2 (R) | negative | 90-1105 | 38-430 | 5367.6 | .015 | angry > neutral |
|  |  |  |  |  |  | happy > neutral |
| 5 (R) | negative | 90-127.5 | 0-758 | 14,130 | .005 | angry > neutral |
|  |  |  |  |  |  | angry > happy |
|  | neutral | 107.5-150 | 0-790 | 13541 | .008 | angry > happy |
|  |  |  |  |  |  | neutral > happy |
| 10 (R) | happy | 90-130 | 280-650 | 5585.5 | .014 | neutral > angry |
|  |  |  |  |  |  | happy > angry |
| 6 (L) | neutral | 135-150 | 156-612 | 5997.7 | .019 | neutral > angry |
|  |  |  |  |  |  | neutral > happy |

*Notes*. Statistical data are depicted only for marginally significant or significant effects of the factor expression. Patients are sorted based on the hemisphere of their amygdala implantations. Abbreviations: L = left, Pat = patient, R = right, TC = target condition.

**Supplementary Table 4**

Significant post-hoc tests for the single-patient analysis (LG)

| Pat | TC | Differences | t | df | p (uncorr.) |
|---|---|---|---|---|---|
| 1 (R) | neutral | angry > neutral | 2.377 | 80 | .019 |
|  |  | happy > neutral | 3.638 | 79 | < .001 |
| 3 (L) | negative | angry > neutral | 4.031 | 62 | < .001 |
|  |  | angry > happy | 4.945 | 60 | < .001 |
|  | neutral | angry > happy | 4.064 | 60 | < .001 |
|  |  | neutral > happy | 2.388 | 49 | .02 |
| 4 (L) | negative | angry > happy | 2.666 | 67 | .009 |
|  |  | neutral > happy | 4.203 | 66 | < .001 |
|  | positive | angry > neutral | 2.536 | 71 | .013 |
|  |  | happy > neutral | 3.668 | 70 | < .001 |
| 6 (L) | positive | happy > angry | 2.513 | 76 | .014 |
|  |  | happy > neutral | 2.337 | 75 | .022 |

*Notes*. Statistical data are depicted only for marginally significant or significant effects of the factor expression. Patients are sorted based on the hemisphere of their amygdala implantations. Abbreviations: Pat. = patient, TC = target condition.

**Supplementary Table 5**

Significant post-hoc tests for the single-patient analysis (HG)

| Pat | TC | Differences | t | df | p (uncorr.) |
|---|---|---|---|---|---|
| 2 (R) | negative | angry > neutral | 3.993 | 72 | < .001 |
|  |  | happy > neutral | 2.371 | 73 | .02 |
| 5 (R) | negative | angry > neutral | 2.887 | 62 | .005 |
|  |  | angry > happy | 3.741 | 60 | < .001 |
|  | neutral | angry > happy | 3.782 | 60 | < .001 |
|  |  | neutral > happy | 1.834 | 49 | 0.072 |
| 10 (R) | happy | neutral > angry | 2.758 | 59 | .007 |

Attention tunes gamma-band responses to faces

| | | | | | |
|---|---|---|---|---|---|
| | | happy > angry | 3.706 | 63 | < .001 |
| 6 (L) | neutral | neutral > angry | 3.043 | 79 | .003 |
| | | neutral > happy | 3.442 | 79 | < .001 |

*Notes*. Statistical data are depicted only for marginally significant or significant effects of the factor expression. Patients are sorted based on the hemisphere of their amygdala implantations. Abbreviations: Pat. = patient, TC = target condition.